\documentclass[pra,aps,preprint,superscriptaddress,amsmath,amssymb,showpacs]{revtex4}

\usepackage{graphicx}
\usepackage{times}
\usepackage{bm}

\newcommand{\rmH}{\mathrm{H}}
\newcommand{\abs}[1]{\left|#1\right|}
\newcommand{\cc}{{\textstyle *}}
\newcommand{\di}{\mathrm{d}}
\newcommand{\what}{\widehat}
\newcommand{\hydm}{{\rmH_2}}
\newcommand{\hydp}{{{\rmH_2}^+}}
\newcommand{\Wpscm}{\mathrm{W}/\mathrm{cm}^2}
\newcommand{\up}{U_\mathrm{p}}
\newcommand{\ip}{I_\mathrm{p}}

\usepackage{color}

\begin{document}
\title{Intrinsic channel closing in strong-field single ionization of $\mathbf{\mathrm{H}_2}$}

\author{Stefan~Pieper}
\affiliation{Max-Planck-Institut f\"ur Kernphysik, Saupfercheckweg 1, 69117 Heidelberg, Germany}
\author{Manfred~Lein}
\affiliation{Institut f\"ur Physik, Universit\"at Kassel, Heinrich-Plett-Stra\ss e 40, 34132 Kassel, Germany}

\date{\today}

\begin{abstract}
The ionization of $\mathrm{H}_2$ in intense laser pulses is studied by numerical integration of the
time-dependent Schr\"odinger equation for a single-active-electron model including the vibrational
motion. The electron kinetic-energy spectra in high-order above-threshold ionization are strongly
dependent on the vibrational quantum number of the created ${\mathrm{H}_2}^+$ ion. For certain vibrational
states, the electron yield in the mid-plateau region is strongly enhanced. The effect is attributed
to channel closings, which were previously observed in atoms by varying the laser intensity.
\end{abstract}


\maketitle
Above-threshold ionization (ATI) \cite{Agostini1979,Eberly1991} of atoms or molecules by intense
laser fields stands for the absorption of more photons than needed to overcome the ionization threshold.
A simple  analysis of classical electron trajectories shows that electrons rescattering once from
the core after the initial ionization step attain large final energies up to $10 \up$
\cite{Paulus1994,Paulus1994a,Paulus1995},
while direct (unscattered) electrons have a maximum energy of $2 \up$. Here, $\up$ denotes
the ponderomotive potential. A striking phenomenon arises when high-order ATI
is studied with respect to its dependence on laser intensity. For the rescattering plateau between $2 \up$
and $10 \up$, it was found in experiment \cite{Hansch1997,Hertlein1997} and calculations
\cite{Muller1998,Muller1999,Paulus2001,Kopold2002,Popruzhenko2002,Wassaf2003,Krajewska2006},
that a slight change in laser intensity can lead to order-of-magnitude changes
in yield for groups of peaks within the plateau. Explanations were found in terms of
multiphoton resonances with Rydberg states \cite{Muller1998,Muller1999,Potvliege2006}
and within the framework of quantum paths \cite{Kopold2000,Paulus2001,Kopold2002}.
The spectral enhancements can be related to channel closings that occur when the (ponderomotively shifted)
lowest ATI peak coincides with an effective threshold \cite{Paulus2001,Kopold2002}.

In the present work, we investigate the role of molecular vibration in ATI channel closings, using the
example of the $\hydm$ molecule. Due to the additional degree of freedom, an additional energy scale
is involved in the dynamics.
In the case of atoms, channel closings were introduced by scanning through a certain range of laser intensities.
We show that for molecules, due to the coupling between electronic and nuclear motion,
\textit{intrinsic} channel closing effects can be observed by  comparing the electron-spectra for different
vibrational states of the created $\hydp$ ion, i.e., applying only one laser intensity.

The minimum energy of a free electron in the presence of a laser field
with amplitude $F_0$ and frequency $\omega$ is equal to
the ponderomotive potential (atomic units are used unless stated otherwise) $\up=F_0^2/4\omega^2$,
which is the quiver energy of the free oscillating electron. Therefore, the laser field modifies
the ionization threshold.
In non-resonant $n$-photon ionization of an atom, the  electron
carries the final kinetic energy
\begin{equation}
  E_\text{kin}=n\,\omega-\ip-\up,\label{eq_ekin-atom}
\end{equation}
with the ionization potential $\ip$  
and integer $n$.
The minimum number $s$ of photons needed to free a bound electron is therefore
defined through $s=\text{ceil}[(\ip+\up)/\omega]$.
By scanning through a range of laser intensities and thus varying $\up$, one can let
ATI peaks disappear at the beginning of the spectrum. Such a channel closing causes
build-up of electron probability near the core \cite{Muller1998} and therefore leads
to significant enhancements of groups of peaks within the rescattering plateau of the ATI spectrum.
Note however, that the precise intensities at which this effect occurs deviate from the estimate
based on the simple formula above \cite{Paulus2001,Kopold2002}.

Dealing with \textit{molecules}, due to the coupling of electronic and vibrational motion, energy is also transferred
to the nuclei of the system, leading to the occupation of vibrationally excited states. We expect that
for a given vibrational state of the created $\hydp$ ion, Eq.~(\ref{eq_ekin-atom}) is changed to
\begin{equation}
  E_\text{kin}=n\,\omega-\ip-\up-\Delta E^v,\label{eq_ekin-mol}
\end{equation}
where $\Delta E^v=E^v-E^0$ is the difference in vibrational energy between
the vibrationally excited state $v$ in question and the vibrational ground state.
Note that $\ip$ denotes here the adiabatic, not the 
vertical ionization potential~\footnote{The adiabatic ionization potential is the 
difference between the total ground state energies of $\hydp$ and $\hydm$ 
(including nuclear motion).}.

Our model of the $\hydm$ molecule consists of a single active electron, interacting with two protons that
are screened by a second (inactive) electron. The electronic
motion is restricted to the polarization direction of
the laser field.
We mention that the effect of moving nuclei in
ATI of 1D $\hydp$ has been studied previously \cite{Bandrauk2003}, but not in the context
of channel closings. 
Note that the dynamics of two active electrons coupled to the 
vibrational motion has been treated earlier within 1D models of $\hydm$
\cite{Kreibich2001,Saugout2007}, but so far it has not been achieved to calculate ATI 
spectra within that approach. In the present work, the molecular 
alignment is perpendicular, i.e.,
the polarization direction is perpendicular to the nuclear motion.
This choice was made
to eliminate the dipole coupling between the electronic ground and first excited state of the
$\hydp$ ion and hence allow for high vibrational excitation of
the electronic ground state \cite{Urbain2004}.
This leads to the Hamiltonian
\begin{equation}
  \begin{split}
  \what{H}(z,R,t) = &-\frac{1}{2}\biggl(\frac{1}{\mu_\mathrm{n}}\frac{\di^2}{\di R^2}
  + \frac{1}{\mu_\mathrm{e}}\frac{\di^2}{\di z^2}\biggr)\\
  &\qquad\qquad+ V_\mathrm{n}(R) + V_\text{int}(z,R) + E(t)z,
  \end{split}
\end{equation}
where the operator $E(t)z$ with $E(t)=F(t)F_0\sin(\omega t)$
describes the interaction of the electron with the electric field of a linearly polarized laser pulse
in length gauge. The quantity $F(t)$ defines
the pulse shape and $F_0$ is the maximum field amplitude. The electron coordinate and inter-nuclear distance
are denoted by $z$ and $R$; $\mu_\mathrm{e}$ and $\mu_\mathrm{n}$ denote the reduced masses
of the active electron and of the two nuclei, respectively.
The vibrational motion in the $\hydp$ ion is incorporated in the model by inserting the exact Born-Oppenheimer
ground-state potential of $\hydp$,
\begin{equation}
  V_\mathrm{n}(R) = V_{\text{BO}}^{\hydp}\!(R).
\end{equation}
This choice reflects the assumption that the second (inactive) electron stays in the ground state
at all times. The active electron interacts with the $\hydp$ ion via a soft-core potential
\begin{equation}
  V_{\text{int}}(z,R) = - \frac{1}{\sqrt{z^2+\sigma^2(R)}}.
\end{equation}
The idea behind this model is that the core, consisting of the proton 
charges screened by the inactive electron, is treated as one, singly 
charged object. The values of $\sigma(R)$
are fitted such that the ground-state Born-Oppenheimer potential of the $\hydm$ model Hamiltonian
matches the exact $\hydm$ Born-Oppenheimer
ground-state potential $V_\text{BO}^\hydm\!(R)$ known from the literature \cite{Kolos1985}.
A similar fitting procedure has been used previously to
reproduce the Born-Oppenheimer potential of $\hydp$ in a 1D model \cite{Feuerstein2003}.
Since our model does not allow excitation of the second electron, we 
have excluded the Coulomb explosion channel, which has been extensively 
studied for example in \cite{Chelkowski2007}.

The propagation of the time-dependent wave function is based on the split-operator method,
along with 2D Fourier transformations to apply the kinetic-energy-dependent operators as
simple multiplications in momentum space. The two-dimensional
grid ($z$-spacing 0.36~a.u., $R$-spacing 0.05~a.u.) extends in $R$-direction from 0.2~a.u.
to 12.95~a.u., in electronic direction from -276.3~a.u. to 276.3~a.u., corresponding to 256 and
1536 grid points, respectively. In the electronic dimension, the grid is further extended up to
$\abs{z}=$ 2522.7~a.u. using a splitting technique \cite{Grobe1999}:
in this outer region the 2D wave function $\Psi_\mathrm{out}(z,R,t)$ is expanded into
products states, i.e., 
\begin{equation}
  \Psi_\mathrm{out}(z,R,t) = \sum_j \xi_j(z,t)\zeta_j(R,t),
\end{equation}
where $\xi_j(z,t)$ and $\zeta_j(R,t)$ are so-called canonical basis states or
natural orbitals \cite{Lodwin1955}, making a single summation over product states possible.
They are obtained as the eigenstates of the one-particle density matrices of the two coordinates $z$
and $R$, respectively, for those portions of the wave function that are transferred to the outer region.
The number of expansion terms is chosen to
keep at least 99.9\% of the total probability. No more than four terms were needed in each
expansion. Within the outer region,
the interaction potential $V_\text{int}(z,R)$ is replaced by the $R$-independent
potential $V_\mathrm{out}(z)=V_\mathrm{int}(z, R=2)$. The coupling between $z$ and $R$
is thus removed in this area, and 1D propagations can be applied separately
to the functions $\xi_j(z,t)$ and $\zeta_j(R,t)$, which allows for large grids. This helps to keep
the entire probability on the grid in spite of the vast electron excursions in strong pulses. It also leads to
high resolution of the kinetic-energy spectra. In the range of $\abs{z}=10\ldots 161$~a.u.,
the interaction is smoothly varied from $V_\mathrm{int}$ to $V_\mathrm{out}$.
This means that unlike \cite{Grobe1999}, our calculation does not involve a discontinuity of the Hamiltonian at the
boundary between inner and outer grid.


The pulse shape $F(t)$ is chosen such that the temporal
pulse integral vanishes. The pulses have a plateau with constant intensity around the middle of the pulse.
The leading and trailing edge of the pulses are $\sin^2$-shaped
ramps. The total pulse length counts five cycles, where 1.5 cycles are used to ramp the pulse on and off,
respectively. The plateau therefore extends over two cycles.
After the end of the pulse,
three empty cycles follow to allow for the continuum electrons to significantly escape from
the Coulomb potential of the $\hydp$ ion before the simulation ends. The laser wavelength is 800~nm
so that we have $\ip=10.0~\omega$.
The system is regarded as ionized for $\abs{z}>$ 30~a.u. The precise choice of this value is not
important since the results shown in this work involve electrons driven at least 500 a.u. from the ion
at the end of the simulation.

Projecting the ionized part $\Psi_\mathrm{ion}$ (see above) of the final wave function onto the
different vibrational states $\chi_v$ of $\hydp$,
\begin{equation}
  \phi_v(z) = \int\chi_v^\cc(R)\,\Psi_\mathrm{ion}(z,R,t_\mathrm{end})\;\di R,
\end{equation}
leads to one ATI spectrum
for each vibrational state considered. The spectra are calculated via Fourier transformation to
$\widetilde\phi_v(p_z)$,
followed by rescaling of the modulus square $|\widetilde{\phi}_v(p_z)|^2$ from momentum to energy.
Only the right-going part of the wave function is used in this work.

According to Eq.~(\ref{eq_ekin-mol}), the ATI peaks within these spectra are
shifted by $\Delta E^v$. Hence, channel closings can be observed by variation of $v$.
Since $\Delta E^v$ can easily exceed the photon energy (for the $\hydp$ ion and $\lambda=800$ nm,
$\Delta E^7\approx 1.0\,\omega$), the channel closing will always take place, no matter
where exactly the first ATI peak is located.
The different vibrational levels play the role of the different laser intensities in the atomic case.
As soon as the vibrational energy eats up enough energy, we expect that the energy spectrum
for the corresponding electrons shows the characteristic channel closing features known from atoms.

\begin{figure}
  \centering
  \includegraphics[width=.85\columnwidth]{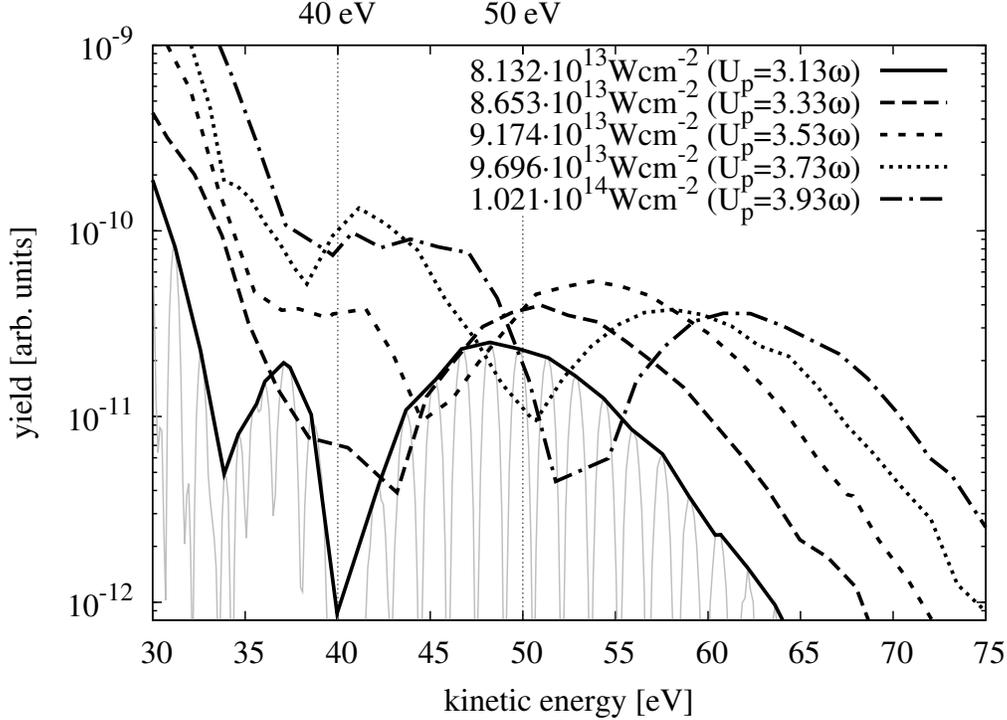}
  \caption{\label{fig_cc}Envelopes of kinetic-energy spectra
    of (right-going) ATI electrons leaving the $\hydp$ ion in the
    $v=4$ vibrationally excited state. The various laser intensities correspond to
    ponderomotive potentials between $\up=3.13~\omega$ and $\up=3.93~\omega$.}
\end{figure}

In Fig.~\ref{fig_cc}, kinetic-energy spectra
of ATI electrons belonging to $\hydp$ ions that
are produced in the $v=4$ vibrationally excited state are shown.
To enhance readability, except for one example only the envelopes are plotted.
Since the spectra correspond to a single vibrational state
of the ion, an atom-like ATI spectrum arises for each laser intensity, and
an atom-like channel closing can be identified. All spectra contain only data from the
$z>0$ part of the grid because a slight difference in peak positions between $z<0$ and $z>0$
would lead to blurring of the peaks. The envelope top reaches its highest
value for an intensity of $I=9.70\times 10^{13}~\Wpscm$ ($\up=3.73~\omega$) at around $42~\mathrm{eV}$.
Trying to estimate this intensity from in Eq.~(\ref{eq_ekin-mol}), the term $\Delta E^v$ is kept constant while
$\up$ is scanned through.
Using Eq.~(\ref{eq_ekin-mol}), we find the channel closing ($E_\mathrm{kin}=0$) to be located
at $\up=(k+0.40)~\omega$, with integer, non-negative $k$, where $k+11=n$,
corresponding to a higher laser intensity of approximately $I = 1.14\times 10^{14}~\Wpscm$
for $n=15$ or a lower one of approximately $I = 8.82\times 10^{13}~\Wpscm$ for $n=14$.
Note  that these values refer to our model Hamiltonian using fitted potentials and ignoring
small effects such as mass polarization.

\begin{figure}
  \centering
  \includegraphics[width=.85\columnwidth]{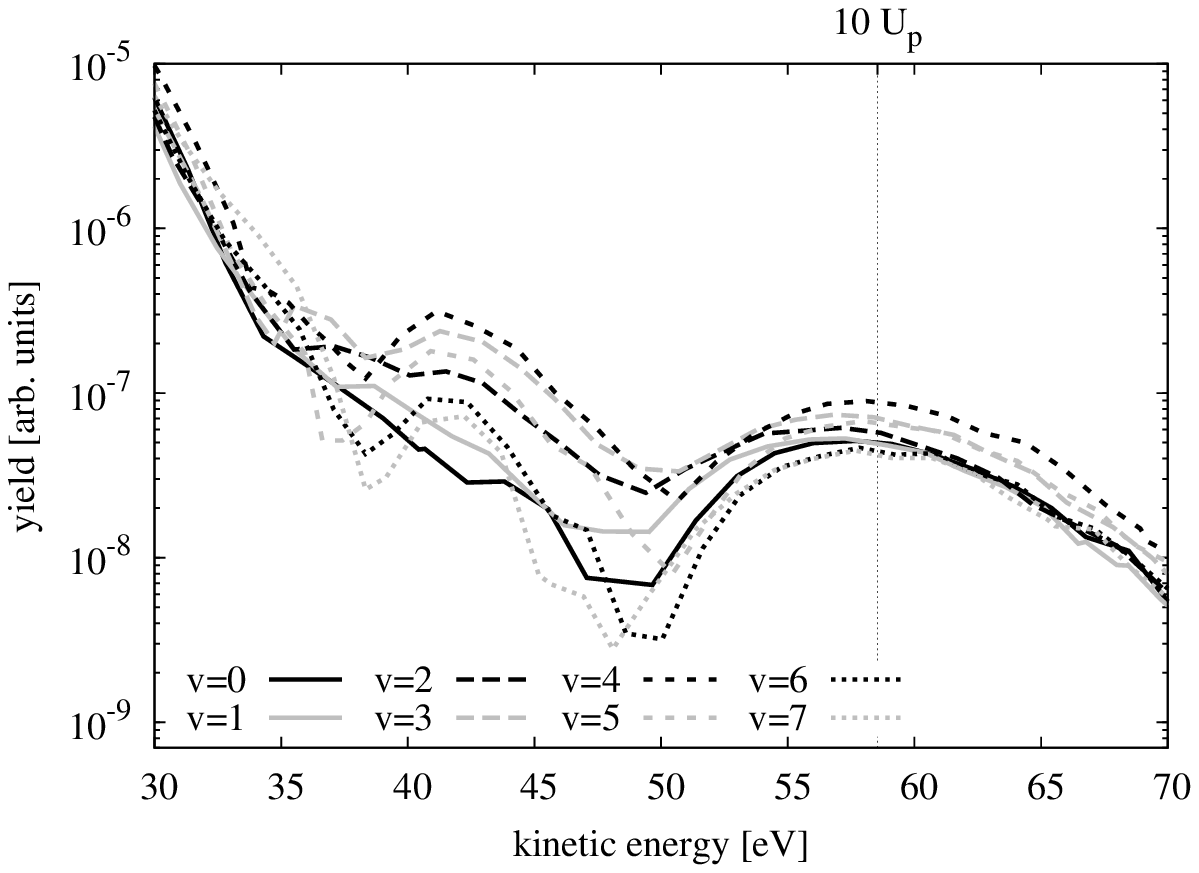}
  \includegraphics[width=.85\columnwidth]{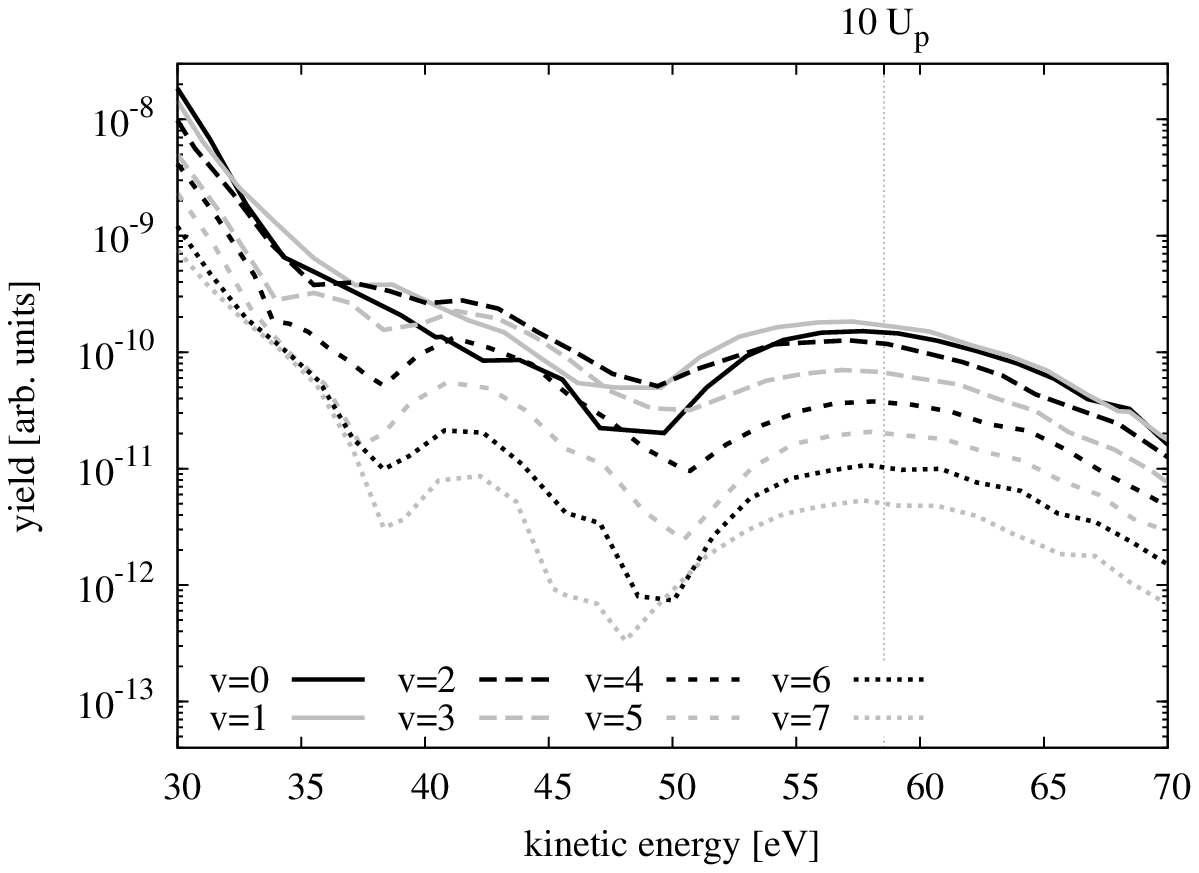}
  \caption{\label{fig_icc}Envelopes of
    kinetic-energy spectra of (right-going) ATI electrons produced by a laser pulse of the intensity
    $I=9.696\times 10^{13}$ $\Wpscm$ ($\up=3.73~\omega$), plotted for vibrational states of the
    $\hydp$ ion from $v=0,\ldots, 7$. In (a), the spectra are divided by the total yield
    of the corresponding vibrational state (cf. Fig.~\ref{fig_vibs}). In (b) the spectra show
    the correct weighting with respect to their yield.}
\end{figure}

\begin{figure}
  \centering
  \includegraphics[width=.85\columnwidth]{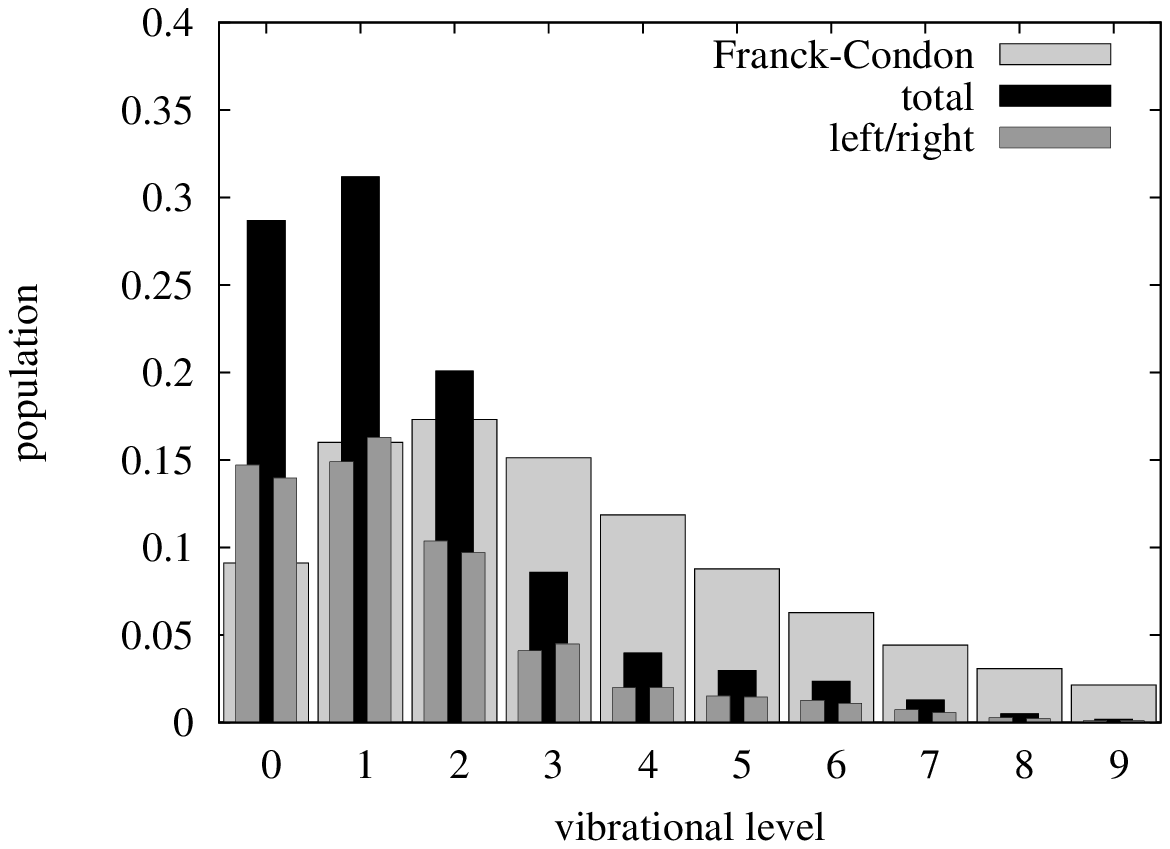}
  \caption{\label{fig_vibs}Occupation of vibrational states after a laser pulse with an
    intensity of $I=9.696\times 10^{13}\Wpscm$ ($\up=3.73~\omega$),
    split up into contributions from the left and right parts of the grid.
    The Franck-Condon overlap is plotted for comparison.
    The distribution has been normalized to total probability one
    .}
\end{figure}

If the laser intensity is kept fixed, but the spectra for several vibrational states of the created ion
are plotted (see Fig.~\ref{fig_icc}), an \textit{intrinsic} channel closing (ICC) appears.
In this case, in Eq.~(\ref{eq_ekin-mol}),
$\up=3.73~\omega$ is kept fixed and $\Delta E^v$ is scanned from $v=0$ to $v=7$.
This corresponds to values between $\Delta E^0=0.00~\omega$ and $\Delta E^7\approx 1.02~\omega$.
The amount of energy the electron loses in each case
as a consequence of energy conservation shifts the spectra with respect to each other.
The ICC is masked by the fact that the
different vibrational states are not populated equally (see Fig.~\ref{fig_vibs}).
Therefore, normalized spectra are plotted in Fig.~\ref{fig_icc}(a),
where each spectrum has been divided by the total yield of the corresponding vibrational state.
The $v=4$ spectrum shows
the highest yield in the middle hump compared to all other spectra within the plot.
We attribute this behavior to an ICC. The unnormalized spectra in Fig.~\ref{fig_icc}(b)
show the highest yield of the middle hump already at $v=2$ due to the suppression of higher
vibrational states.
The application of Eq.~(\ref{eq_ekin-mol}) leads to $\Delta E^v=(k+0.31)~\omega$, where
$k+14=n$, which
corresponds to an energy difference close to the vibrational state $v=2$ (using $n=14$).
Again,
the observed position of the channel closing is shifted with respect to the one expected from
Eq.~(\ref{eq_ekin-mol}).

It should be
stressed that the energy difference between two vibrational states is larger than $0.1~\omega$ for all
vibrational states considered in this work.
Hence, in contrast to the intensity scanning,
the transition over a channel closing is not sampled continuously.
Similar calculations for $\mathrm{D}_2$ are work in progress and provide a finer graining due
to the closer-lying vibrational states of the ${\mathrm{D}_2}^+$ ion.
\begin{figure}
  \centering
  \includegraphics[width=.85\columnwidth]{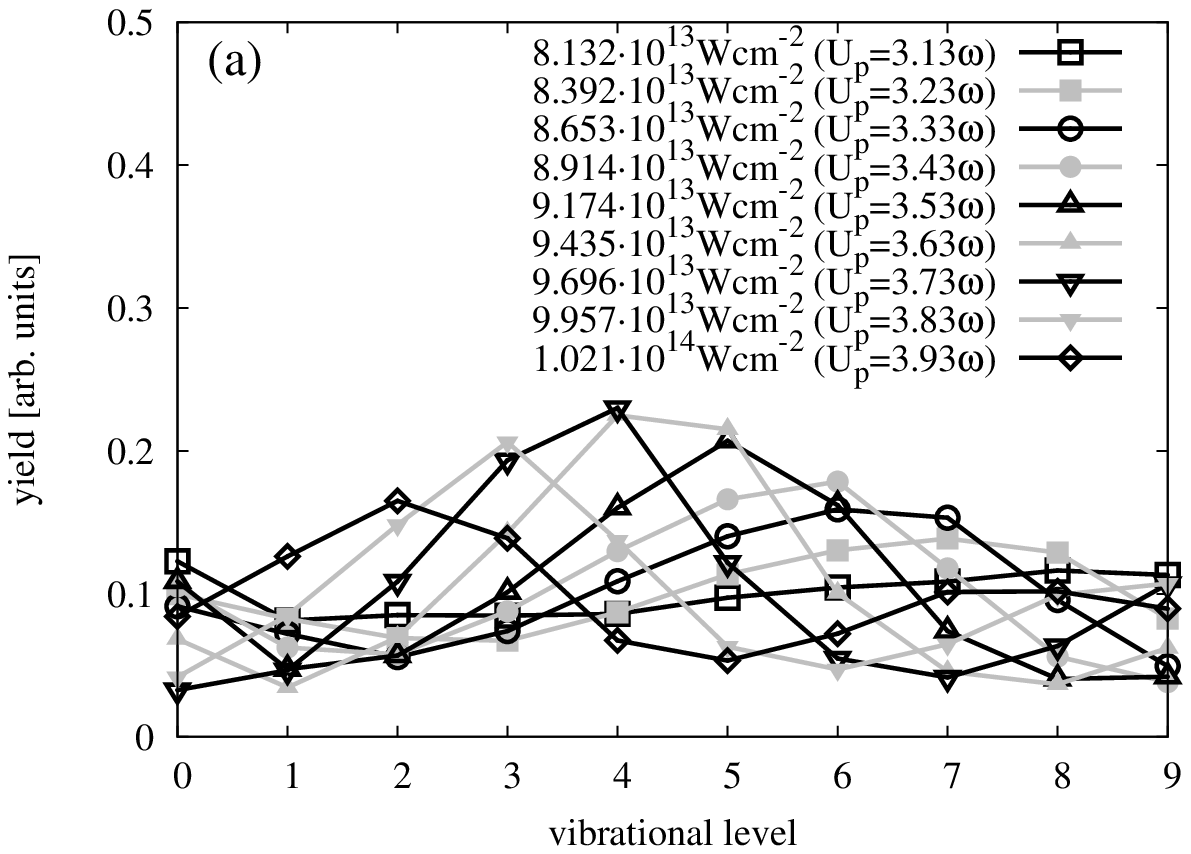}
  \includegraphics[width=.85\columnwidth]{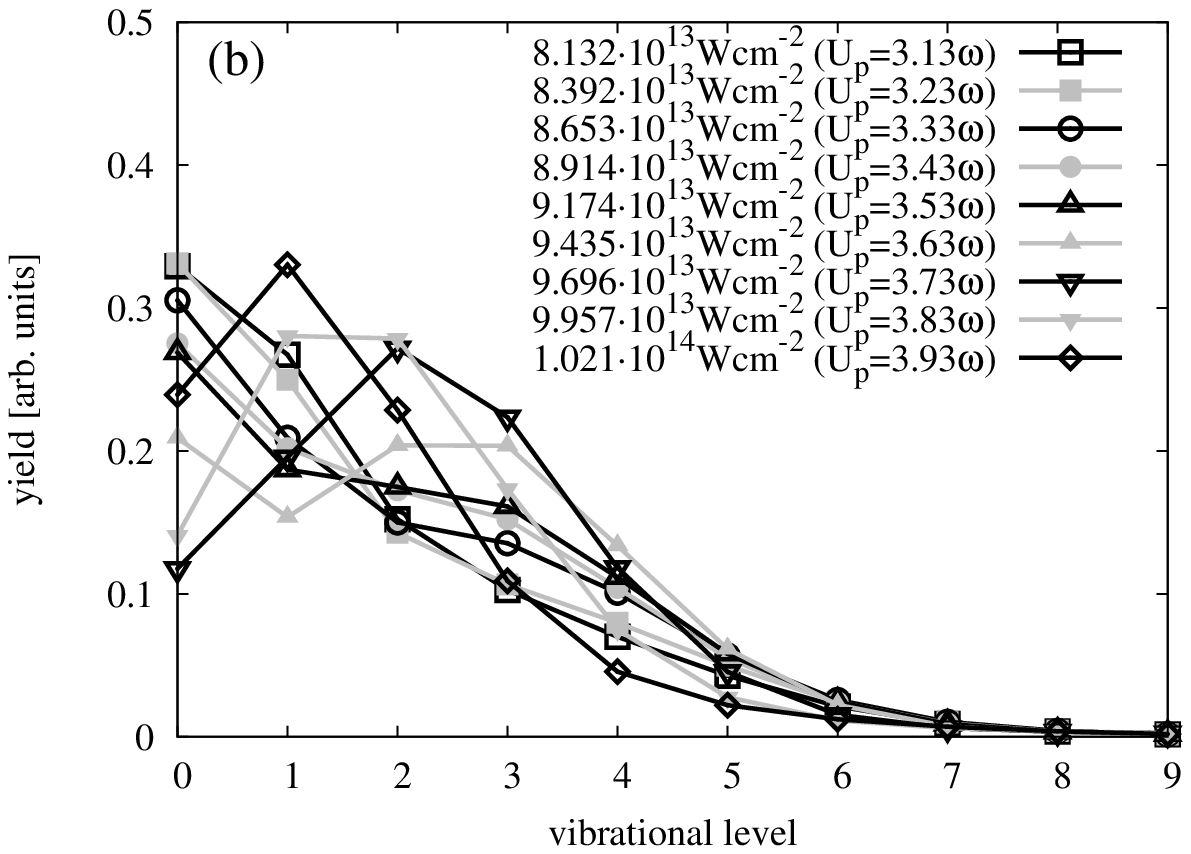}
  \caption{\label{fig_e_over_vib}Electronic yield within
    the energy window $40$-$50$~eV (cf. Fig.~\ref{fig_cc}), plotted versus vibrational states of the $\hydp$ ion
    for different laser intensities. In (a), the electron spectra used were divided by the total yield
    of the corresponding vibrational state (cf. Fig.~\ref{fig_vibs}) as in Fig.~\ref{fig_icc}(a).
    In (b), correctly weighted spectra with respect to their yield were taken as in Fig.~\ref{fig_icc}(b).}
\end{figure}

We show in Fig.~\ref{fig_e_over_vib}, that for a suitable electron-energy window
the ICC structure appears in the electron yield plotted as a function of vibrational quantum numbers. The
distributions are shown using either the normalized electron yield from Fig.~\ref{fig_icc}(a), see
Fig.~\ref{fig_e_over_vib}(a), or the unnormalized electron yield from Fig.~\ref{fig_icc}(b), see
Fig.~\ref{fig_e_over_vib}(b).
We use the energy window between 40 and 50~eV, corresponding to the dashed lines
in Fig.~\ref{fig_cc}.
Clearly the ICC shows up in Fig.~\ref{fig_e_over_vib}(a).
Although the ICC feature is not as evident in the unnormalized distributions of Fig.~\ref{fig_e_over_vib}(b),
it is clearly visible that electrons and ions are highly correlated, since the distribution
over vibrational states is, within the chosen electron-energy window, very different from the general distribution
of vibrational states for all electrons. See, e.g., the curve for $\up=3.73~\omega$ as compared to
Fig.~\ref{fig_vibs}.

To summarize, we found clear signatures of spectral enhancements due to channel closings occurring
in ATI of $\hydm$ by scanning through the vibrational states of the created $\hydp$ ion.
The explanation of this effect seems straightforward, applying energy conservation
to the photon-absorbing molecule. Similar to atoms, we find that the effect occurs at intensities/vibrational
states slightly different from the simple estimate based on the unperturbed ionization potential.
We conclude with a note on the experimental perspectives. The populations of $\hydp$ vibrational states
after strong-field ionization of $\hydm$ has been measured in \cite{Urbain2004}, but a measurement
in coincidence with electrons will be difficult. On the other hand, coincidence measurements similar to
recent pump-probe experiments \cite{Ergler2006} appear feasible. The goal would be to measure the electron from an
ionizing pump pulse, together with fragments from probe-pulse-induced Coulomb explosion of $\hydp$.

This work has been supported by the Deutsche Forschungsgemeinschaft.

\end{document}